\documentclass[letterpaper, preprint, paper,11pt]{AAS}	% for preprint proceedings

\usepackage{bm}
\usepackage{amsmath}
\usepackage{subfigure}
\usepackage[colorlinks=true, pdfstartview=FitV, linkcolor=black, citecolor= black, urlcolor= black]{hyperref}
\usepackage{overcite}
\usepackage{footnpag}			      	% make footnote symbols restart on each page

\PaperNumber{XX-XXX}

\begin{document}

\title{Proximity Operations about Apophis through its 2029 Earth Flyby}

\author{Daniel J. Scheeres\thanks{Distinguished Professor, Smead Department of Aerospace Engineering Sciences, University of Colorado, Boulder, Colorado, 80309, USA. \\ Presented as Paper IAA-AAS-SciTech-034 at the 2nd IAA/AAS SciTech Forum, Moscow, Russia, June 2019.}
}

\maketitle{}

\begin{abstract}
The dynamics and control of a satellite in proximity to the asteroid Apophis across its Earth close approach in 2029 is evaluated and investigated. First, the feasibility of carrying out close proximity operations about Apophis when in its heliocentric orbit phase is evaluated and shown to be feasible. Then three different types of close proximity motion relative to Apophis are analyzed that will enable a spacecraft to take observations throughout the Earth close approach. These are maintaining a relative orbit that is somewhat distant from Apophis, hovering along the Earth-Apophis line, or maintaining orbit about Apophis through the flyby. Each of these are shown to be feasible, albeit challenging, and some basic aspects of these operations are noted and discussed. 
\end{abstract}

\section{Introduction}

The 2029 flyby of Earth by the asteroid (99942) Apophis will be a spectacle for all humanity to observe. The asteroid will be close enough to the Earth to be visible during its close approach, at approximately 37,200 km from the center of the Earth (under 6 Earth radii). A number of proposed missions are in development for taking advantage of its close Earth passage in order to measure what effects the strong Earth tidal forces may have as it passes through closest approach \cite{ivashkin2012optimal,ivashkin2016analysis,plescia2017asteroid}. These concepts include having both landed and orbital elements about this small asteroid. Previous analyses have shown that the surface forces and changes will be modest, even though the rotation state will change significantly, and thus that landed elements may be feasible \cite{scheeres2005abrupt,yu2014numerical,demartini2019using}. This paper will instead consider the relative dynamics of any co-orbiting vehicles about Apophis during its close approach to Earth, in order to evaluate if it will be feasible to both stay in close proximity to the asteroid during the Earth closest approach, and what level of control effort may be required to enable spacecraft relative observations through the entire close approach passage. Previous analysis has looked at the feasibility of orbiting about Apophis \cite{ivashkin2016analysis,ivashkin2017analysis}, however they have not considered the feasibility of maintaining orbit or proximity through the closest approach to Earth. This analysis uses the recently measured Apophis shape and spin state based on radar measurements \cite{brozovic2018goldstone}. 

%99942 Apophis          	
%Discovered 2004 June 19 by R. A. Tucker, D. J. Tholen and F. Bernardi at Kitt Peak.
%Also known as Apep, the Destroyer, Apophis is the Egyptian god of evil and destruction who dwelled in eternal darkness. As a result of its passage within 40~000 km of the earth on 2029 Apr. 13, this minor planet will move from the Aten to the Apollo class.

This analysis will look at a number of different approaches for maintaining proximity through the Earth flyby. These include having a spacecraft in the vicinity of the asteroid (but not in orbit about it), a spacecraft in orbit about the asteroid, and a spacecraft actively hovering in close proximity to the asteroid. For some of the proposed scientific investigations it will be crucial that a spacecraft in proximity observe the asteroid throughout the entire closest approach phase. The challenge is that the spacecraft will be perturbed by the relative dynamics induced by the flyby, which has a closest approach of 37,200 km and a hyperbolic eccentricity of 4.232. Thus there may be challenges to maintaining a useful relative orientation to the body. By studying the effects of the flyby on different relative orbits it will be possible to better design any candidate mission to this body. 

The paper is structured as follows. First we review the model of Apophis, including its spin state and shape in addition to its orbital characteristics. Next we introduce the different approaches for maintaining observation of Apophis before, during and after the closest approach. We study the placement of a spacecraft in orbit about Apophis and in a neighboring heliocentric orbit in particular, showing that both of these approaches are feasible. Finally, we discuss the implications of our results and state our conclusions. 

\section{Model of Apophis and its Earth Flyby}

\begin{figure}[htb]
	\centering\includegraphics[width=5in]{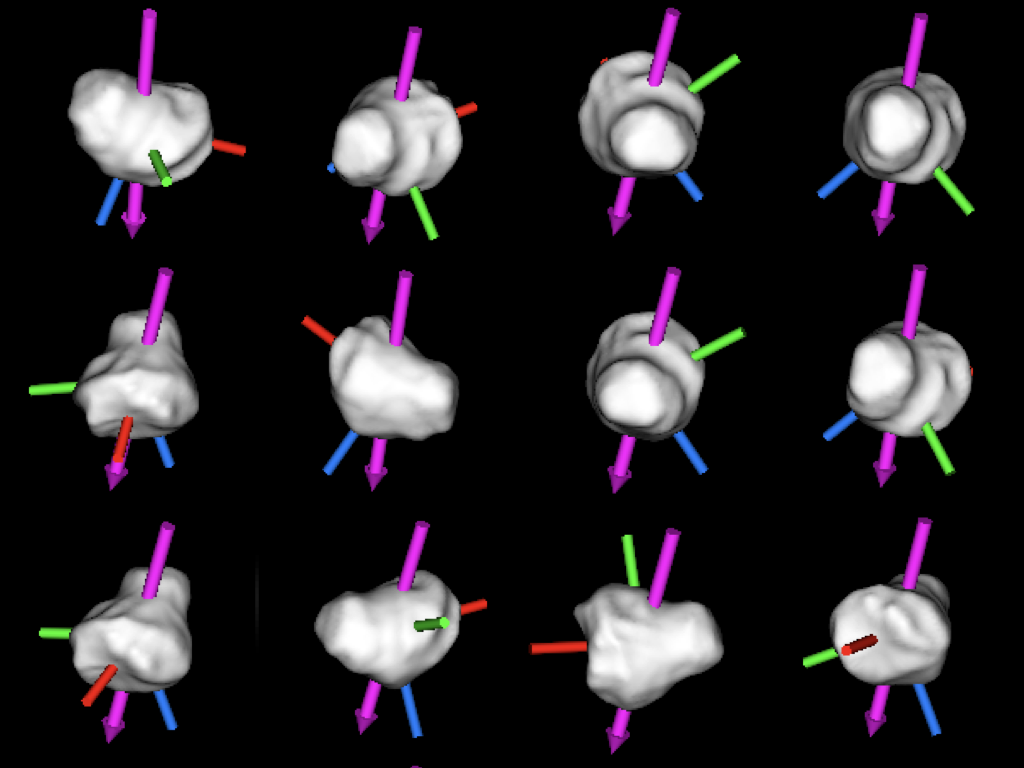}
	\caption{Views of the Apophis radar shape model in its tumbling rotation state \cite{brozovic2018goldstone}. }
	\label{fig:apophis}
\end{figure}

Since its discovery the Apophis system has been observed photometrically and using radar. Both of these observation campaigns have developed models for the asteroid shape and spin state which agree well with each other \cite{pravec2014tumbling,brozovic2018goldstone}. In this paper we use the radar shape model and its associated values for for the shape as well as the associated model. Figure \ref{fig:apophis} shows several views of the radar shape model in its tumbling rotation state. Table \ref{tab:parameters} lists some of the important physical parameters derived from this model. Spectroscopic observations of the asteroid have determined that it is in the S-Type complex\cite{binzel2009spectral}, consistent with a bulk density of 2 g/cm$^3$, which is what we assume herein. Apophis has been documented to lie in a complex rotation state \cite{pravec2014tumbling,brozovic2018goldstone}, presumably due to past close passages to Earth \cite{scheeres2005abrupt}. Detailed spin state definitions are given in both of these documents, however for the current study we just model the excitation without trying to precisely match the expected orientation at Earth's closes approach in 2029 (this due to likely uncertainties). The asteroid is in a Short-Axis Model (SAM) spin state, meaning that it is relatively close to rotation about its maximum moment of inertia. The solution does place it near the intermediate axis, however, signifying a high degree of excitation. Figure \ref{fig:apophis} shows the shape following this complex rotation. The purple vector signifies the angular momentum, as seen from the Earth around its close passage in 2013. The colored axes are fixed in the body frame along its principal axes and allow one to track its rotational motion.

\clearpage

\begin{table}[h]
\caption{Basic physical parameters for Apophis \cite{brozovic2018goldstone}.}
\label{tab:parameters}
\centering
\begin{tabular}{| c | c | c | c |}
\hline
Parameter & Value & Units & Notes \\
\hline
\hline
 Mean Radius		& 	168	& 	m	& Geometric Mean \\
\hline
 Largest Extent		& 	410	& 	m	& \\
\hline
 Volume		& 	$1.986\times10^7$	& 	m$^3$	& \\
\hline
 Gravitational Parameter $\mu_A$		& 	2.650	& 	m$^3$/s$^2$	& Assumes density of 2 g/cm$^3$ \\
\hline
 Moment of Inertia Ratios $I_{xx}/I_{zz}$		& 	0.7294	& 	--	& Constant density \\
\hline
 Moment of Inertia Ratios $I_{yy}/I_{zz}$		& 	0.9479	& 	--	& Constant density \\
\hline
 Spin State $I_D$		& 	0.9752	& 	--	& SAM rotation state \\
\hline
 Spin Period $P_{l}$		& 	30.6	& 	h	& Equivalent rotation period\footnote{This is the rotation period of a sphere with the same total angular momentum and a moment of inertia $I_D$.} \\
\hline
 Angular Momentum Latitude		& 	-59.3	& 	deg	& Earth Ecliptic Frame \\
\hline
 Angular Momentum Longitude		& 	246.8	& 	deg	& Earth Ecliptic Frame \\
\hline
\end{tabular}
\end{table}

\footnotetext{This is the rotation period of a sphere with the same total angular momentum and a moment of inertia $I_D$.} 

What makes Apophis of particular interest is its orbit, which takes it quite close to the Earth on April 13, 2029. Its heliocentric and geocentric orbit parameters are listed in Table \ref{tab:orbit}. For the Earth flyby, we mapped its geocentric orbit elements into a frame defined by the asteroid rotational angular momentum. \textcolor{black}{To do this we chose the z-axis along the angular momentum vector, the x-axis along the Earth-Apophis line at closest approach, and the y-axis following the right-hand rule.} This is relevant for the design of orbits about Apophis, as in general they will be designed relative to the body frame, however how the Earth influences these orbits will be significant as well. Although we note that the Earth flyby will change the Apophis rotation state and angular momentum pole, we do not include that level of detail in the current analysis. 

\begin{table}[h]
\caption{Basic orbit parameters for Apophis \cite{brozovic2018goldstone}.}
\label{tab:orbit}
\centering
\begin{tabular}{| c | c | c | c |}
\hline
Parameter & Value & Units & Notes \\
\hline
\hline
\multicolumn{4}{|c|}{Heliocentric Elliptic Orbit}\\
\hline
 Semi-Major Axis		& 	0.9224	& 	AU	&  \\
\hline
 Eccentricity		& 	0.1912	& 	--	& \\
\hline
 Periapsis Radius		& 	0.746	& 	AU	& \\
\hline
 Apoapsis Radius		& 	1.099	& 	AU	& \\
\hline
 Inclination		& 	3.3	& 	deg	& Earth Ecliptic \\
\hline
 Argument of Periapsis		& 	126.4	& 	deg	&  Earth Ecliptic \\
\hline
 Longitude of Asc. Node		& 	204.4	& 	deg	&  Earth Ecliptic \\
 \hline
 \hline
\multicolumn{4}{|c|}{Geocentric Hyperbolic Orbit}\\
 \hline
 Periapsis Radius	($q$)	& 	$3.72\times10^4$	& 	km	&  \\
\hline
 Eccentricity		& 	4.229	& 	--	& \\
\hline
 Inclination		& 	47.8	& 	deg	& Relative to Apophis Angular Momentum \\
\hline
 Argument of Periapsis		& 	-143.9	& 	deg	&  Relative to Apophis Angular Momentum\\
\hline
 Longitude of Asc. Node		& 	1.8	& 	deg	& Relative to Apophis Angular Momentum \\
\hline
 Time of Periapsis Passage		& 	2029-Apr-13 21:46:04.8	& 	--	&  \\
\hline
\end{tabular}
\end{table}

\clearpage

\section{Close Proximity Operations during the Heliocentric Phase}

Before we investigate the dynamics of a spacecraft in close proximity to Apophis during the Earth flyby, it is important to first establish what the feasible operations are for a spacecraft about Apophis when it is not in close proximity to the Earth. There are three basic approaches we will consider, defined as follows. Distant: defined as being in a close proximity orbit to Apophis, yet far enough away so that the Apophis attraction is not relevant. Close Hovering: defined as being close enough to Apophis to require the gravity field to be accounted for, and actively thrusting to maintain a given distance from the body. Orbiting: defined as being in an orbit about Apophis. Each of these can be analyzed in detail, however in this paper we only provide a brief summary of results based on more detailed analyses \cite{scheeres_asteroid_book, scheeres_smallbody, broschart_hovering_1, broschart_hovering_2}. 

\subsection{Distant Orbit Operations}

A satellite can maintain itself on a neighboring orbit relative to Apophis' heliocentric orbit, yet far enough so that the gravitational attraction of Apophis can be neglected.  In general there is a rich class of relative motions that could be investigated, discussed in part in \cite{scheeres_china}. However, the effect of solar radiation pressure must be accounted for when in such a relative orbit. The simplest motion is a relative equilibria on the line between the asteroid and the sun, located at a distance of $X = -\beta/3 d$ where $d$ is the Apophis-Sun distance which changes in orbit, and 
\begin{eqnarray}
	\beta & = & \frac{(1+\sigma) P_o}{\mu_{S} B}
\end{eqnarray}
where $\sigma$ is the reflectance of the spacecraft, $P_o \sim 1\times 10^8$  kg km$^3$/ s$^2$ / m$^2$ is the solar constant which accounts for the photon momentum transport, $\mu_S \sim 1.327 \times 10^{11}$ km$^3$/s$^2$ is the Sun's gravitational parameter and $B$ is the spacecraft's mass to area ratio, a value which can range from $10\rightarrow 100$ kg/m$^2$. Many interplanetary spacecraft have a mass to area ratio between $25-50$, while a 1 kg cubesat would have a value of 100. 

For Apophis, the parameter $\beta = 7.54\times10^{-4} / B$ (assuming a near-zero $\sigma$) and the distance $d$ varies between $1.12\rightarrow1.65 \times 10^{8}$ km over its heliocentric orbit. Thus, the distance of this solar radiation pressure equilibrium from the asteroid will range between $8.4 \times 10^3$ km at perihelion and $1.24 \times 10^4$ km at aphelion for $B=10$, and will be cut down at most by an additional factor of 10 for a small and heavy spacecraft. These distances are all far away enough from the asteroid so that the neglect of its gravitational attraction is appropriate. For characterization purposes we will assume a value of $B=50$ and evaluate the situation at 1 AU, around the Earth closest approach. Then the SRP equilibrium distance is 2300 km. 

This is too distant for detailed observation with most low-cost telescopes, thus it would be desired to place the spacecraft closer to the asteroid. This can be achieved by implementing a hovering mode, although here the asteroid gravity is not being actively nulled, but instead the tidal and solar radiation pressure effects are being adjusted. This equilibrium distance can be moved in to be closer to the asteroid through the use of thrusting along the Sun-Asteroid line. Applying a thrust of $A$ towards the sun will in general move the equilibrium a distance $\frac{d^4}{3 \mu_S a_H(1-e_H^2)} A$ towards the asteroid, where $d$ is the sun-asteroid distance, $\mu_S$ is the Sun's GM, $a_H$ is the Apophis heliocentric orbit semi-major axis and $e_H$ is the Apophis heliocentric orbit eccentricity. At 1 AU, to move the equilibrium point 2000 km towards the asteroid will require an acceleration on the order of $2\times10^{-7}$ m/ s$^2$, or 0.017 m/s/day, which is quite modest. Due to these very small accelerations, the spacecraft can be placed much closer to the asteroid and not undergo any significant migration over a time span of days. 
These distances are too far away from Apophis to be realistically maintained through the flyby, however. Thus, if a distant operations approach is taken it is likely that the spacecraft will have to be repositioned to be closer to Apophis leading up to the flyby. 

\subsection{Close Hovering Operations}

If the spacecraft is brought within a few km of the asteroid surface, then the gravitational attraction of the asteroid must be taken into account. This is another way in which a spacecraft can position itself in such a way to have a close view of the asteroid before, during and after the Earth flyby. This begins to mimic the operations approach used by the Hayabusa and Hayabusa2 missions \cite{hirabayashi2022hayabusa2}. \textcolor{black}{A useful metric for characterizing the control effort needed to hover is to evaluate the body's gravitational attraction at a given distance, and multiplying by total time.} For Apophis, at a distance of 1 km it will take a thrust of about $3\times10^{-6}$ m/s$^2$ to maintain its position, which integrates to about 0.26 m/s/day. In general, a spacecraft's position can be maintained at an arbitrary relative position to the asteroid. Although this mode of hovering is naturally unstable, the technology to stabilize and maintain it have been developed and studied in detail previously \cite{broschart_hovering_2, scheeres_asteroid_book}. 

\subsection{Orbit Mechanics about Apophis}

\begin{figure}[htb]
	\centering\includegraphics[width=5in]{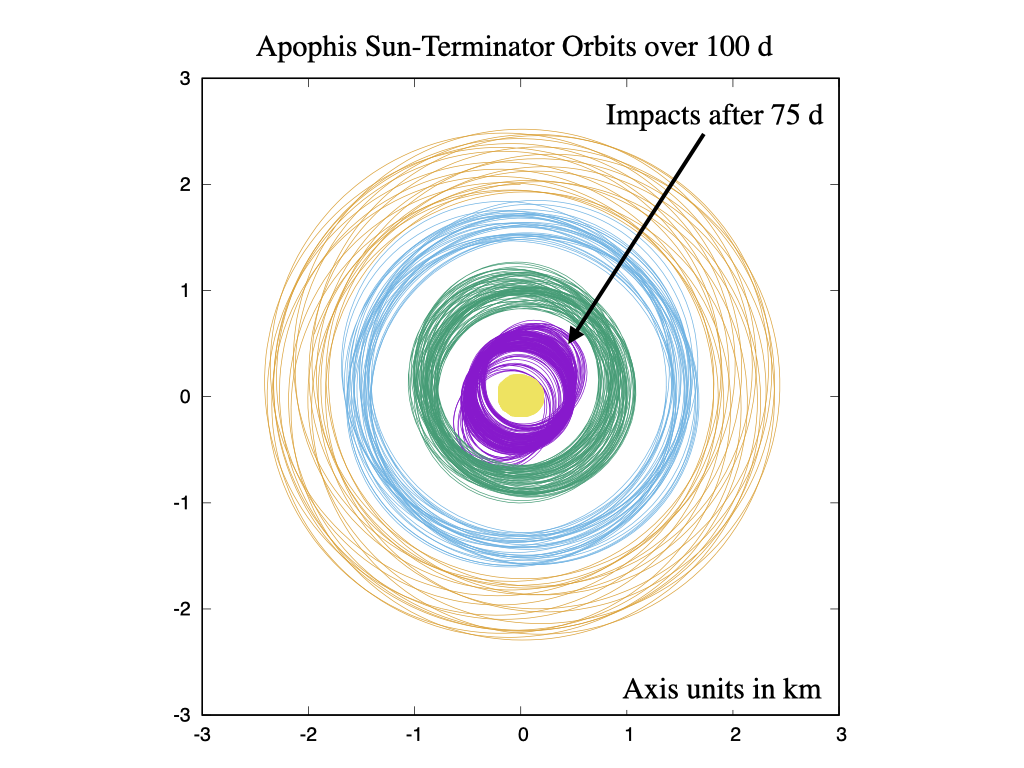}
	\caption{Initially circular terminator orbits about Apophis as viewed from the sun. The innermost (initial semi-major axis of 0.5 km) impacts after 75 days. The others all remain in orbit. }
	\label{fig:helio_terminator}
\end{figure}

When visiting Apophis in its heliocentric orbit phase, there are specific limits on the design of spacecraft orbits to be stable in the Apophis regime. The major perturbations are from the non-spherical gravity field of Apophis and from solar radiation pressure (SRP). Since Apophis has a low overall spin period and a complex spin state, the impacts of the gravity field are  significant and can lead to resonances between the spacecraft orbital motion and the Apophis rotational motion \cite{hu_slowrotate}. Despite this, the perturbations from SRP are dominant. Borrowing from the analysis in \cite{scheeres_asteroid_book, scheeres_smallbody}, the maximum limit on semi-major axis for a body to remain in orbit can be derived to be 
\begin{eqnarray}
	a_{Max} & = & 0.3344 \sqrt{B} d
\end{eqnarray}
which gives the maximum semi-major axis in units of km, while $B$ is the mass to area ratio and will have values from $10\rightarrow100$ kg/m$^2$ and $d$ is the Sun-Apophis distance in AU and will range from $0.75\rightarrow 1.1$. Thus, at a distance of 1 AU a spacecraft with $B=50$ must have an orbit that has semi-major axis less than 2.4 km. There is concern if the maximum semi-major axis is close to the radius of the body, as gravitational distributions can be coupled with SRP to create a chaotic environment. However here we see that the semi-major axis as measured in radii is $2 \sqrt{B} d$, and for our example is greater than 10 radii at the lower limit, which occurs at perihelion. 
Thus, placing a satellite in orbit about Apophis is definitely feasible. The strength of the SRP perturbation would suggest that a terminator orbit design would be appropriate for an Apophis orbiter \cite{scheeres_smallbody}. \textcolor{black}{A terminator orbit is an orbit situated in the plane perpendicular to the sun-line. When orbiting a small body, such as Apophis, these orbits will naturally track the sun, due to the SRP perturbation acting on the spacecraft. In this regime, placing a satellite in a plane significantly out of the terminator generally results in the spacecraft impacting on the asteroid, due to secular growth of the eccentricity.} Figure \ref{fig:helio_terminator} shows some example orbits \textcolor{black}{about the Apophis shape model gravity field, accounting for the effect of SRP and the asteroid's orbit about the sun. These orbits are} started as circular orbits in the sun-terminator orbit plane. It is possible to specify initial conditions that have much less variation in their orbit, however we do not focus on these questions here. We do note that an initial orbit with semi-major axis of 0.5 km is destabilized and impacts in about 75 days. This is due to resonant interactions with the complex spin state of the asteroid. Lower orbits will impact more rapidly, and thus are not considered in this paper. 

\section{Close Proximity Operations during the Earth Flyby Phase}

The main focus of this paper follows, and considers different possible dynamical operations that will keep the spacecraft in close proximity to Apophis during Earth flyby. This is an important aspect for any mission which wishes to observe and track that asteroid through its close passage with the Earth. This is a challenging dynamical regime, however, thus motivating our study. 

For our detailed dynamical computations we will focus on a total time period of 25 days around the closest approach, starting about 4.5 days prior to closest approach and continuing in orbit for another 20.5 days. \textcolor{black}{We note that the main effect of the Earth on the orbit occurs over just a few hours around closest approach. However, we choose this longer time span to first verify that the spacecraft orbit is stable prior to the flyby, and to gauge the level of instability of the orbit after the flyby.} During this period we do not consider the perturbation from the sun, although this assumption must be reconsidered later should any of these proximity dynamical models be pursued. We first outline the basic dynamics equations used for our study and then analyze different operational modes through flyby. 

\subsection{Proximity Dynamics Equations of Motion}

We consider all of our dynamics in a frame that rotates with the Apophis-Earth line, with angular rate given by the relative hyperbolic orbit. Thus, the angular rate of this frame is changing in its orbit and, at an instantaneous value of true anomaly, its rate and its acceleration equal
\begin{eqnarray}
	\dot{f} & = &  \sqrt{\frac{\mu_E}{q^3(1+e)^3}} (1+e\cos f)^2 \\
	\ddot{f} & = & - \frac{2\mu_E}{q^3(1+e)^3} e \sin f
\end{eqnarray}
\textcolor{black}{where $\mu_E$ is the gravitational parameter of the Earth and $q$ is the periapsis distance of the Apophis flyby hyperbola from the center of the Earth, computed from the semi-major axis and eccentricity as $q=a(1-e)$. }
We see that the magnitude of $\sqrt{ \mu_E / q^3(1+e)^3} \sim 7.4\times10^{-6}$ rad/s controls the strength of the kinematic terms. 

These quantities are then present in the scalar equations of motion for a particle relative to the Earth. These equations are derived in detail in \cite{scheeres_asteroid_book}, and the $x$ axis points from the Earth to Apophis, the $z$ axis is along the geocentric orbit angular momentum and the $y$ axis completes the triad. 
\begin{eqnarray}
	\ddot{x} - 2 \dot{f} \dot{y} - \ddot{f} y - \dot{f}^2 x & = & \frac{\partial V}{\partial x} + 2 \frac{\mu_E}{d^3} x + A_x \nonumber \\
	\ddot{y} + 2 \dot{f} \dot{x} + \ddot{f} x - \dot{f}^2 y & = &  \frac{\partial V}{\partial y} - \frac{\mu_E}{d^3} y + A_y  \label{eq:full} \\
	\ddot{z} & = &  \frac{\partial V}{\partial z} - \frac{\mu_E}{d^3} z + A_z \nonumber 
\end{eqnarray}
The quantity $V$ is the Apophis gravitational force potential. For our detailed numerical simulations we use the potential corresponding to the constant density Apophis shape model, however for analytical estimates we use the point mass approximation $\mu_A / r$.  The quantitiy $d$ is the distance of Apophis from the Earth and equals $d = q(1+e) / (1+e \cos f)$. These equations incorporate the Hill approximation for the gravitational attraction of the Earth relative to the Apophis center of mass. Finally, the quantities $A_x$, $A_y$ and $A_z$ are control accelerations that the spacecraft can implement. 

For some of our analytical estimates it is simpler to restate the equations of motion in a pulsating frame with true anomaly as the independent parameter. This is then the hyperbolic Hill three body problem. \textcolor{black}{This is not a typical frame for analysis, although it is fundamentally related to previously derived equations for the elliptic Hill three body problem \cite{scheeres_asteroid_book}. } Performing a scaling by the distance $d$ and changing the independent parameter from time to true anomaly using the transformation $\dot{X} = X' df / dt$, where $X' = d X/ df$, yields the equations 
\begin{eqnarray}
	X'' - 2Y' & = & 
		\frac{1}{1+e\cos f}\left[ 3 - \frac{\mu_A/\mu_E}{R^3}\right] X + \alpha_X \nonumber \\
	Y'' + 2X' & = & 
		- \frac{\mu_A/\mu_E}{R^3} Y + \alpha_Y \label{eq:scaled} \\
	Z'' + Z & = & 
		- \frac{\mu_A/\mu_E}{R^3} Z + \alpha_Z  \nonumber 
\end{eqnarray}
where $R = \sqrt{X^2 + Y^2 + Z^2}$. 
In this form of the equations we just use the point-mass Apophis attraction, although this is not a requirement. Note, we do not make the typical ``final'' scaling of positions by $(\mu_A/\mu_E)^{1/3}$, which would make these equations parameterless except for the eccentricity -- such a complete normalization is not needed here. 
The ratio $\mu_A / \mu_E \sim 6.6\times10^{-15}$, which is quite small. The parameter $\alpha$ is a normalized acceleration which will be analyzed later. For a given acceleration $A$ it is defined as $\alpha = A \frac{d^3}{q(1+e) \mu_E}$. One advantage of this form of the equations is that the complicated kinematic terms all cancel nicely, leaving just a few terms. The disadvantage is that solutions of these equations must be translated back into dimensional space to evaluate their feasibility, this is done by multiplying by the Earth-Apophis distance $d$, or $x = d \ X$, etc.  

\subsection{Distant Monitoring Dynamics}

During the flyby a spacecraft can position itself relative to Apophis and observe the asteroid through the closest approach. If a position is chosen that is relatively far from the asteroid, the gravitational attraction of the asteroid can be neglected. 
The placement of the spacecraft can be controlled such that it passes through the Earth-Apophis direction during the flyby.  
It \textcolor{black}{can be surmised } that this orientation may be of the most benefit for observing possible changes in the asteroid, although other vantage points \textcolor{black}{can} also be considered \textcolor{black}{with the derived equations and results}. 

If the ratio $\mu_A / \mu_E$ is ignored in Eqns.\ \ref{eq:scaled} they simplify to the Tschauner-Hempel (TH) equations, which represent linearized motion about a Keplerian orbit \cite{carter1990new,yamanaka2002new,dang2017solutions}. 
The TH equations can be solved in closed form, with the results given in the Appendix. A spacecraft's trajectory avoiding close approaches to the central asteroid would follow these equations. These are usually used for elliptic motion ($e<1$), however they also apply to motion relative to a hyperbolic orbit for $e>1$. Here, the asteroid is at the origin, which is also an equilibrium point for the system. These equations have been normalized by the Earth-Asteroid distance, $d = \frac{p}{1+e\cos f}$, and have true anomaly as the independent parameter. Thus, given a solution to these, they must be scaled by the Earth-Apophis distance and re-parameterized in time. 

For the solution to the TH equations it is instructive if we combine all terms that contain the drift term $L$ (defined in the Appendix), which increases secularly in time and leads to divergence between two neighboring trajectories. In general we would like to eliminate this term on average. To do this we combine the different solution components that contain the drift term\textcolor{black}{, leaving off additional terms which can be found in the Appendix}. 
\begin{eqnarray}
	X & = & - 3 e \sin f (1+e\cos f) \left[ (2+e)X_o + (1+e) Y_o' \right] L + \ldots \\
	Y & = & -3 (1+e\cos f)^2\left[ (2+e) X_o + (1+e) Y_o' \right] L + \ldots \\
	X' & = & - 3 e ( \cos f + e\cos2f ) \left[ (2+e) X_o + (1+e) Y_o' \right] L + \ldots \\
	Y' & = & 6  e \sin f (1+e\cos f) \left[ (2+e) X_o  + (1+e) Y_o' \right] L + \ldots 
\end{eqnarray}

It is simple to see that all of the drift terms involve the combination of initial conditions $(2+e)X_o + (1+e)Y_o'$, and thus choosing this combination to be zero will ensure that the drift terms do not appear. The simplest way to ensure this is to choose
\begin{eqnarray}
	Y_o' & = & - \frac{2+e}{1+e} X_o 
\end{eqnarray}
Doing this, we also note that the linear momentum integral for the $Y$ equation becomes $H = Y_o' + 2X_o = \left[ 2 - \frac{2+e}{1+e}\right] X_o = \frac{e}{1+e}X_o$. Thus, for the circular case this momentum must be zero, but for the eccentric and hyperbolic cases it is non-zero. 

Substituting these initial conditions into the solutions for $X$ and $Y$ we find
\begin{eqnarray}
	X & = & \frac{1+e\cos f}{1+e}\left\{  \cos f  X_o + \sin f X_o' \right\}\\
	Y & = & Y_o - \frac{1}{1+e}\left\{\sin f(2+e\cos f) X_o - \left[ \cos f(2+e\cos f) - (2+e)\right] X_o' \right\} \\
	Z & = & \cos f Z_o + \sin f Z_o'
\end{eqnarray}
To get a better sense of the geometry of motion, we can put these equations into the general form of an ellipse. Doing so then yields the parametric equation
\begin{eqnarray}
%	\left( X + \frac{2e^2}{1-e^2}X_o\right)^2 + \left(\frac{1+e\cos f}{2+e\cos f}\right)^2 \left(Y-Y_o + \frac{2+e}{1+e}X_o'\right)^2 & = & \left( \frac{1+e\cos f}{1+e}\right)^2 \left[ X_o^2 + X_o'^2\right] \\
	1 & = & \frac{(1+e)^2}{X_o^2 + X_o'^2} \left\{ \left( \frac{X }{1+e\cos f}\right)^2 + \left( \frac{Y-Y_o + \frac{2+e}{1+e}X_o'}{2+e\cos f}\right)^2 \right\}
\end{eqnarray}
From this equation we see that the path of the spacecraft will follow an ``osculating'' ellipse that is centered at $X=0$ and $Y=Y_o-\frac{2+e}{1+e}X_o'$ with a pulsating ``size'' along the $X$ axis equal to $\frac{1+e\cos f}{1+e}\sqrt{X_o^2 + X_o'^2}$, and along the $Y$ axis equal to $\frac{2+e\cos f}{1+e}\sqrt{X_o^2 + X_o'^2}$.

\begin{figure}[htb]
	\centering\includegraphics[width=5in]{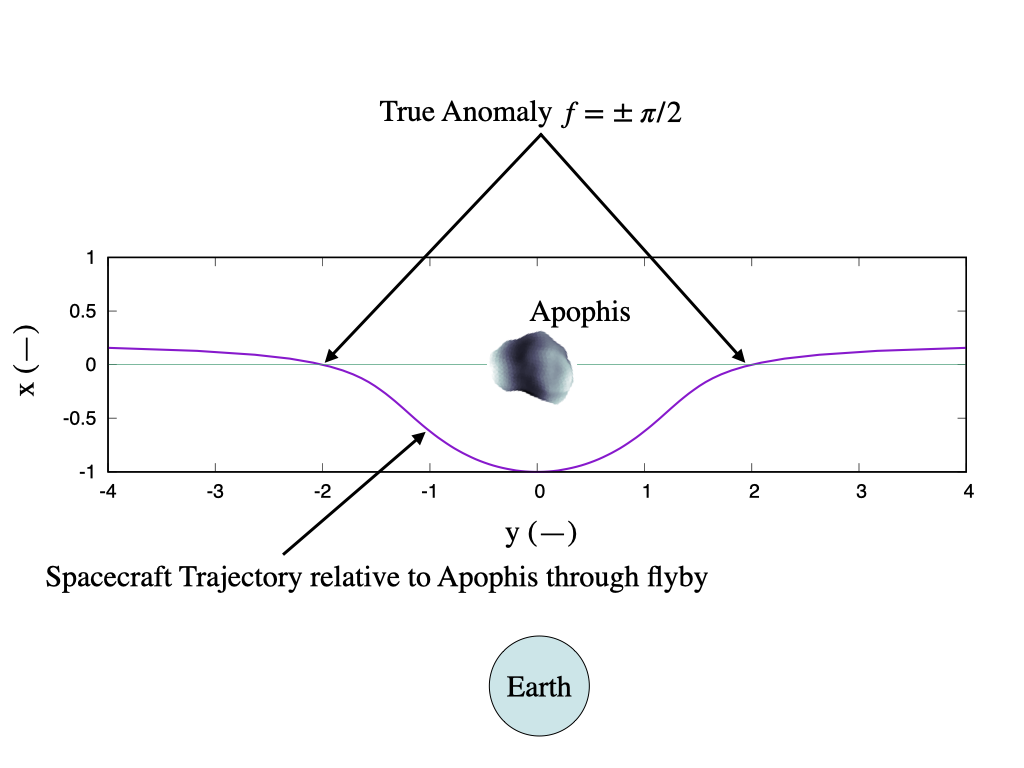}
	\caption{Trajectory of a spacecraft on a neighboring hyperbola with the same drift rate. The conditions are chosen to place the spacecraft along the Apophis-Earth line at closest approach. \textcolor{black}{The plot is in the rotating Earth-Apophis frame.}}
	\label{fig:flyby}
\end{figure}

It is instructive to rescale these by multiplying by the Earth-Apophis distance, $d = \frac{q(1+e)}{1+e\cos f}$,  to find dimensional distances. This yields 
\begin{eqnarray}
	x & = & q \left\{ \cos f  X_o + \sin f  X_o' \right\}\\
	y & = & d Y_o - q \left\{\sin f \frac{2+e\cos f}{1+e\cos f}  X_o - \frac{ \cos f(2+e\cos f) - (2+e)}{1+e\cos f} X_o' \right\} \\
	z & = & d \left[ \cos f Z_o + \sin f Z_o' \right]
\end{eqnarray}
where $q$ is the periapsis radius. A particular approach may be to choose initial conditions such that at closest approach to the Earth the spacecraft lies along the Earth-Apophis line with no velocity component along this line. This corresponds to setting $X_o' = Z_o = Z_o' = 0$. The solution for this motion for a unit offset is shown in Fig.\ \ref{fig:flyby} \textcolor{black}{in the frame rotating with the Earth-Apophis line}. 

An important observation for this mode of close proximity operations is that the natural motion of the satellite is to transition from one side of the asteroid to the other, passing through closest approach along the asteroid-Earth line. This also implies that it may be more difficult to maintain a fixed location along the asteroid-Earth line, although this is analyzed next. 

\subsection{Hovering Dynamics}

Another option for a distant flyby is to actively thrust the spacecraft to keep a given distance from the asteroid through the flyby. In this way it is possible to place the spacecraft along the asteroid-Earth line. We will analyze two different approaches, the first keeping the scaled distance constant (meaning that the actual distance will vary during the close approach). The other keeping the actual distance constant. 

\subsubsection{Scaled Hovering Distance}

To keep the scaled distance constant we choose the acceleration $\alpha_x$ to fix a given offset distance $X_H$ in Eqns.\ \ref{eq:scaled}. The resulting hovering distance will then vary in time as $x_H = \frac{q(1+e)}{1+e\cos f} X_H$. The necessary scaled accelerations are 
\begin{eqnarray}
	\alpha_X & = & -\left[3-(\mu_A/\mu_E)/|X_H|^3\right] X_H/(1+e\cos f), 
\end{eqnarray}
which varies with true anomaly, and $\alpha_Y = \alpha_Z = 0$.  The required acceleration is 
\begin{eqnarray}
	A_x = \frac{- \left[3-(\mu_A/\mu_E)/|X_H|^3\right] \mu_E }{q^2(1+e)^2}(1+e\cos f)^2 X_H, 
\end{eqnarray}
and $A_y = A_z = 0$. Note, this equation is generally valid and accounts for the gravitational attraction of Apophis, although this will be minimal unless we come down to a few km from the surface. 

We note that if the distance is chosen such that $X_H = \left(\frac{\mu_A}{3 \mu_E}\right)^{1/3}$ then the required acceleration is zero, as the spacecraft will be located at the Hill equilibrium. If $X_H$ is greater than this in magnitude and between the Earth and Apophis, the net acceleration is positive. Conversely, if $X_H$ is between this equilibrium and Apophis it flips to negative, as then the asteroid gravity will exceed the tide. For the scaled system this distance is $\pm 1.3\times10^{-5}$, and at closest approach corresponds to 0.484 km while at $f=\pm90^\circ$ will be approximately 2.5 km. 

For another example, let us specify a distance of 10 km on the Earth side at closest approach, or $x_H(f=0) = -10$ km. This requires $X_H = -2.7\times10^{-4}$. The hovering spacecraft will then move ``naturally'' from a distance of 52.3 km at $f = \pm90^\circ$ to 10 km at closest approach, with its nominal motion along the Earth-asteroid line. 
The acceleration required to maintain this distance at closest approach then is approximately $2.3\times10^{-4}$ m/s$^2$, and at $f = \pm 90^\circ$ reduces to $8\times10^{-6}$ m/s$^2$. These levels of acceleration are quite modest, and in general will decrease as the hovering point is moved even closer to the asteroid. 

This example is illustrative, although the resulting large variation in motion relative to Apophis is not of particular interest and presents challenges for the variations in surface resolution over a relatively short time period. Thus, we do not recommend this as a useful approach. It is instructive to note that this approach only requires modulating the acceleration along the Earth-Apophis line, with cross accelerations all being zero. 

\subsubsection{Fixed Hovering Distance}

More feasible from a spacecraft observation perspective is to fix a distance from the asteroid to the spacecraft using a hovering approach, and then adjust the hovering acceleration during the Earth flyby to maintain a constant distance. To analyze this it is best to consider the initial equations of motion in a non-scaled system, given in Eqns.\ \ref{eq:full}. Then, for a given hovering location along the Earth-Apophis line, $x_H < 0$, the necessary control accelerations are 
%When far from the Earth, the acceleration needed to hover along the Earth-Apophis line is approximately $A_x = \frac{\mu_A}{x^2}$. For a distance of 500 - 1000 m this cost is 10.6 - 2.65 $\mu$m/s$^2$, or 92 - 23 cm/s/day. During and through the closest approach this total cost will vary appreciably. The correction term is found by balancing the equations of motion. We choose $y = z = 0$ and $x = x_H$ as our nominal hovering point. We note that to maintain our location along the Earth-Apophis line now requires a thrust in the $\hat{\bf y}$ direction -- for the distant hovering dynamics this was not needed due to the naturally changing distance between the spacecraft and Apophis. The thrusting equations can be stated as
\begin{eqnarray}
	A_x & = & \left[ \frac{\mu_A}{|x_H|^3}  - \frac{\mu_E}{q^3(1+e)^3} (1+e\cos f)^4 - 2 \frac{\mu_E}{q^3(1+e)^3}(1+e\cos f)^3 \right] x_H \\
	A_y & = & - \frac{2\mu_E}{q^3(1+e)^3} e \sin f x_H \\
	A_z & = & 0
\end{eqnarray}
As we now fix the location without allowing scaling of the distance, a lateral acceleration $A_y$ is also required. 
The lateral acceleration is maximized at $f = \pm 90^\circ$ and equals $A_y = \mp 4.6\times10^{-10} x_H$, and for a hovering distances of 1-10 km equals $0.46\rightarrow4.6 \times10^{-6}$ m/s$^2$, which integrates to 0.04-0.4 m/s/day, which is modest. 

For the radial acceleration, we first consider the general cost of hovering at a distance $x_H$, which is $A_x = \mu_A / x_H^2$, and for hovering distances from 1-10 km will equal $2.6\rightarrow0.026 \times10^{-6}$ m/s$^2$, which integrates to $0.23 \rightarrow 0.0023$ m/s/day. We note that these values are comparable to the lateral accelerations, but must be corrected to account for the tidal accelerations. 
The radial acceleration correction has its greatest value at perigee, and yields a correction to the hovering term equal to $\left[ 1 - \frac{(\mu_E / \mu_A) |x_H|^3}{q^3}(3+e)\right] \sim 1 - 20 |x_H|^3$ where $x_H$ is measured in km. Thus, for hovering distances from 1-10 km the correction term will vary from $-20\rightarrow-2\times10^4$, meaning that it will in general change direction and vary by orders of magnitude potentially.  The corresponding acceleration magnitude at closest approach is then $52\rightarrow 520 \times10^{-6}$ m/s$^2$. 
For comparison, at a true anomaly of $f = \pm 90^\circ$ the correction terms will be $1 - 6.1\times 10^{-2} |x_H|^3$, or range between $0.94\rightarrow -60$ for our range of hovering distances from 1-10 km. For the lower hovering location the correction required is small, however for the 10 km hovering distance the cost is still an order of magnitude larger and in the opposite direction. 

Thus, even though the overall acceleration magnitudes are small, maintaining a fixed hovering altitude will require careful planning and control of the spacecraft. Given estimates of the change in magnitude it may be feasible for the spacecraft to implement a feed-forward control law that uses the predicted motion in conjunction with current measurements to maintain control. Such a study would be of interest to evaluate whether a closed loop control, such as developed in \cite{broschart_hovering_2}, could adequately control a hovering spacecraft. 

\subsection{Apophis Orbiting Dynamics}

Finally, we consider the feasibility of maintaining an orbit about Apophis during the close Earth passage. There are two issues that must be accounted for in this scenario. First, whether the flyby will perturb the orbit so much that it will impact on the asteroid. Second is whether the orbiter will be stripped out of orbit of Apophis during the flyby. 

First we consider constraints for the orbiter to be stripped out of orbit around Apophis. The nominal conservative limits for this, which have been developed previously in the literature, do not apply well here. The Earth-Apophis equilibrium point will be at $x_E = \pm \left(\frac{\mu}{3\mu_E}\right)^{1/3} q \sim \pm 0.484$ km. This is above the Apophis surface, implying that material will not be shed from the body. However, it is quite close as well. The usual conservative limit for not being stripped out over long orbit periods is that $a < x_E / 3$, i.e., within one-third of the Hill radius, however that is below the Apophis surface. Such analyses do not necessarily apply, however, as the flyby is very rapid relative to orbital motion about Apophis. Thus, the flyby acts as an ``impulse'' applied to the satellite. This is analyzed by taking the ratio of the angular rate of the hyperbola at closes approach to the mean motion angular rate of the spacecraft. This yields $\frac{\dot{f}}{n} = \sqrt{1+e}\left(\frac{a}{x_E}\right)^{3/2}$. If we choose $a \sim x_E$, then the ratio is $\sim 2.2$, meaning that the effect of the tidal flyby will occur rapidly relative to the motion of the spacecraft. This does not mean that this is a minor effect, however, as since the spacecraft will in general be near to being outside of the Hill sphere of Apophis at closest approach it is definitely subject to very large perturbations and may be stripped out of orbit. 

In fact, the situation is so extreme that we must analyze it numerically to see if orbits are possible to continue through closest approach. They are, but are very sensitive to the location of the satellite in its orbit and to the satellite orbital plane. 	To carry out this analysis we initialize different orbits about Apophis, placing the spacecraft in a circular orbit at a range of orbit plane orientations \textcolor{black}{relative to the Apophis total angular momentum}. 

In the following we start our numerical integrations at about 4.5 days before closest approach and propagate until 20.5 days after closest approach to evaluate the orbital stability. 
The simulations account for the shape-derived gravity field of Apophis, its complex spin state and the tidal effect of the Earth during the hyperbolic flyby. For this phase they do not account for the solar tide or solar radiation pressure, and also neglect the change in complex rotation state around closest approach. 
We limit our initial conditions to be circular orbits with a semi-major axis of 0.5 km. A larger semi-major axis will always be stripped out of orbit during the flyby, as it is well outside of the Hill sphere. A lower semi-major axis was seen to be unstable in general, while a 0.5 km orbit was seen to persist for at least tens of days.  
For this set of conditions we study three different orbit planes with a set of limited numerical integrations. These are direct, polar and retrograde orbits. 

\begin{figure}[htb]
	\centering\includegraphics[width=5in]{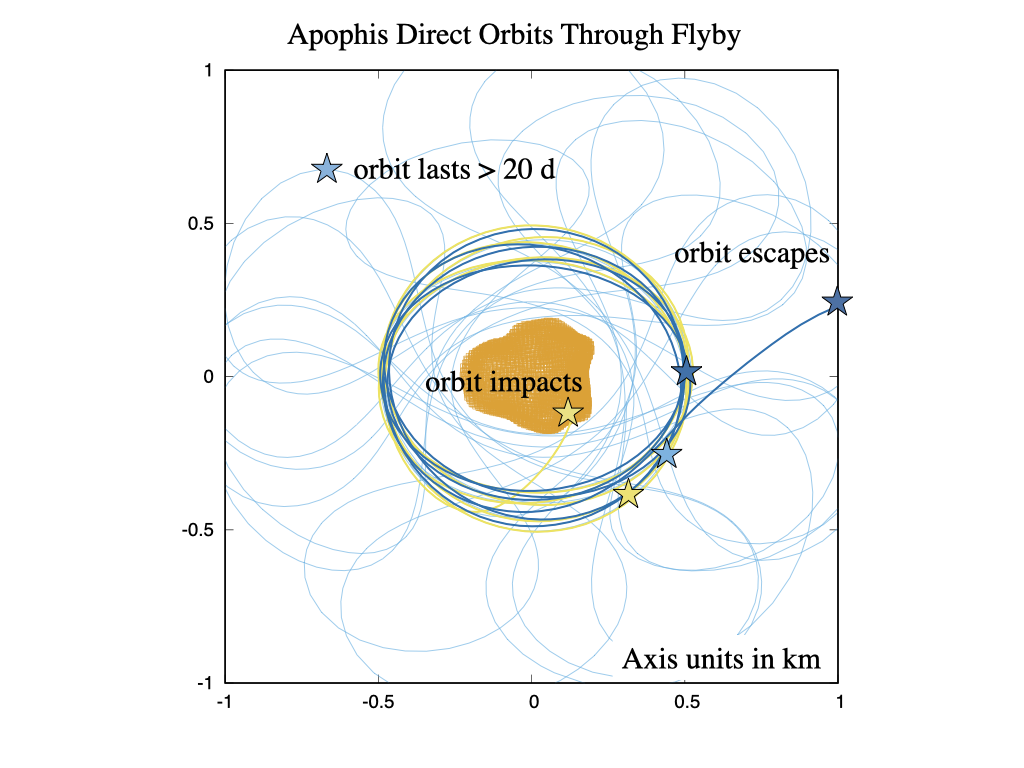}
	\caption{Diverse outcomes of three initially direct orbits following flyby. One escapes, one remains in orbit and one impacts, all starting within 45 degrees of each other. Axis units are in km. }
	\label{fig:direct_orbit}
\end{figure}

First, consider direct orbits with zero inclination relative to the plane perpendicular to Apophis' rotational angular momentum. This configuration is expected to be the most sensitive to perturbations during the flyby, and indeed we find that almost all orbits are stripped out of orbit and escape. However, by phasing the initial conditions of the spacecraft in orbit it is possible to find some orbits that impact and a few that remain tenuously bound in orbit. Figure \ref{fig:direct_orbit} shows three such orbits, with their initial spacing along the circular orbit spanning 45$^\circ$. \textcolor{black}{However, with such extreme sensitivity to spacecraft placement in orbit, it is likely that such direct orbits would be difficult to navigate due to the expected propagation of orbit uncertainties.}

\begin{figure}[htb]
	\centering\includegraphics[width=5in]{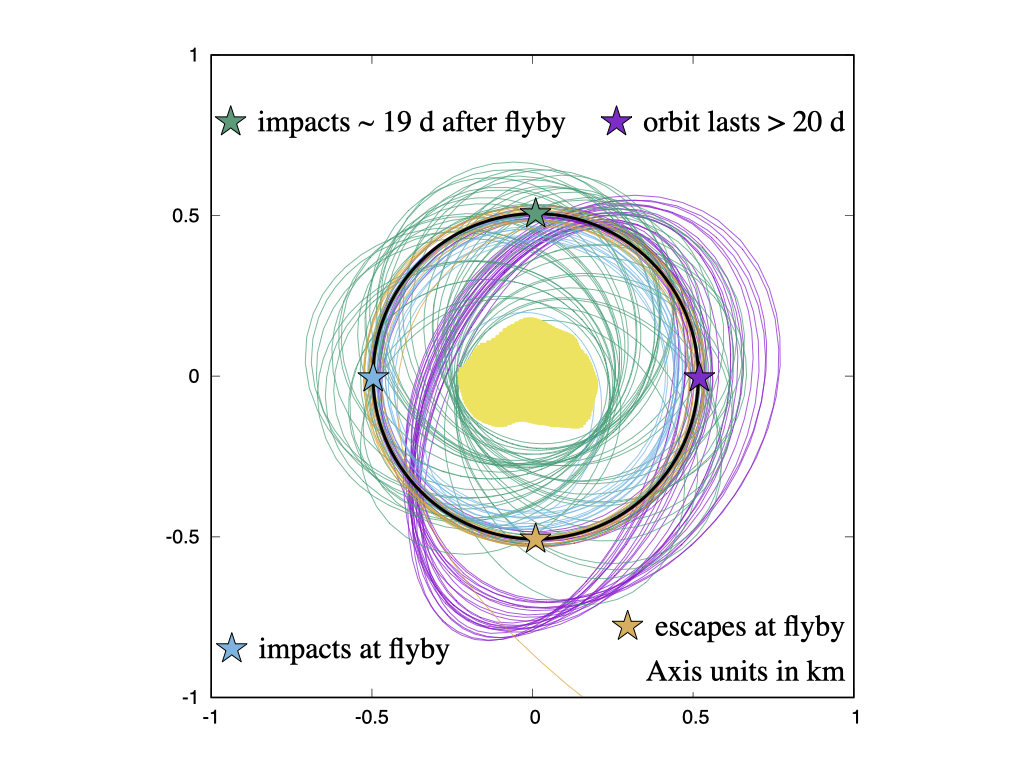}
	\caption{Four initially polar orbits with initial conditions shifted by 90 degrees in true anomaly, as indicated. One remains in orbit more than 20 days after flyby, one impacts just before 20 days, one impacts right after flyby and one escapes right after flyby. Axis units are in km. }
	\label{fig:polar_side}
\end{figure}

\begin{figure}[htb]
	\centering\includegraphics[width=5in]{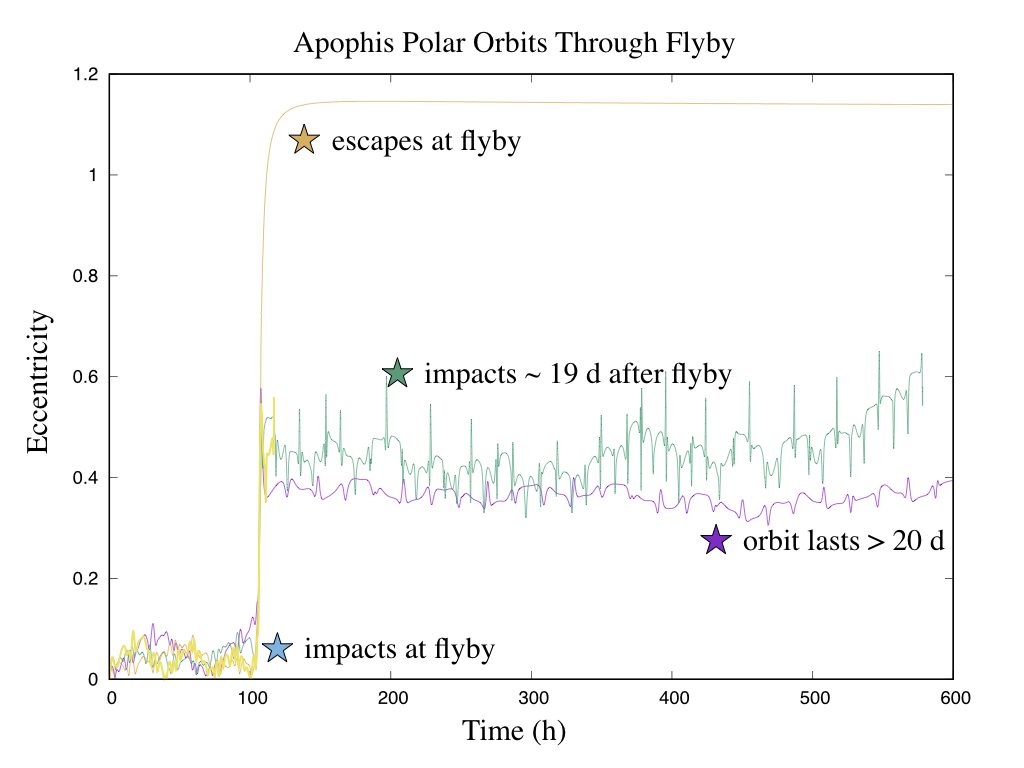}
	\caption{Time history of the eccentricities of the four polar orbits shown in Fig.\ \ref{fig:polar_side}. }
	\label{fig:polar_eccen}
\end{figure}

\begin{figure}[htb]
	\centering\includegraphics[width=5in]{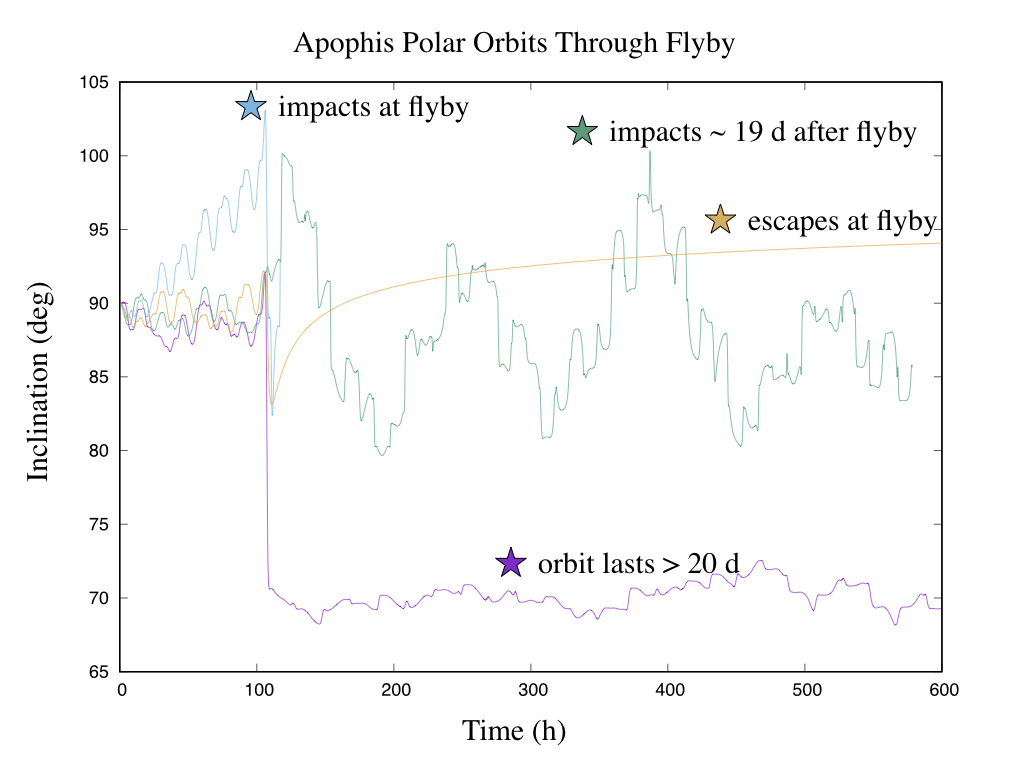}
	\caption{Time history of the inclinations of the four polar orbits shown in Fig.\ \ref{fig:polar_side}. }
	\label{fig:polar_inc}
\end{figure}

Next we consider polar orbits oriented at 90$^\circ$ to the plane perpendicular to the Apophis rotational angular momentum. Here we fix the initial node but vary the location of the spacecraft by 90$^\circ$ steps and see extreme sensitivity to the spacecraft phasing to the eventual outcome. This occurs again as the satellites relative geometry to the Earth at closest approach controls the strength and direction of the tidal perturbation. The resultant orbits range from remaining bound to impacting to escaping again. Figure \ref{fig:polar_side} shows these cases. Figures \ref{fig:polar_eccen} and \ref{fig:polar_inc} provide more insight into the outcomes in terms of orbit elements of eccentricity and inclination. \textcolor{black}{These orbits are better candidates than the direct orbits, as there are apparently larger ranges of placement that can yield bound orbits. }

\begin{figure}[htb]
	\centering\includegraphics[width=5in]{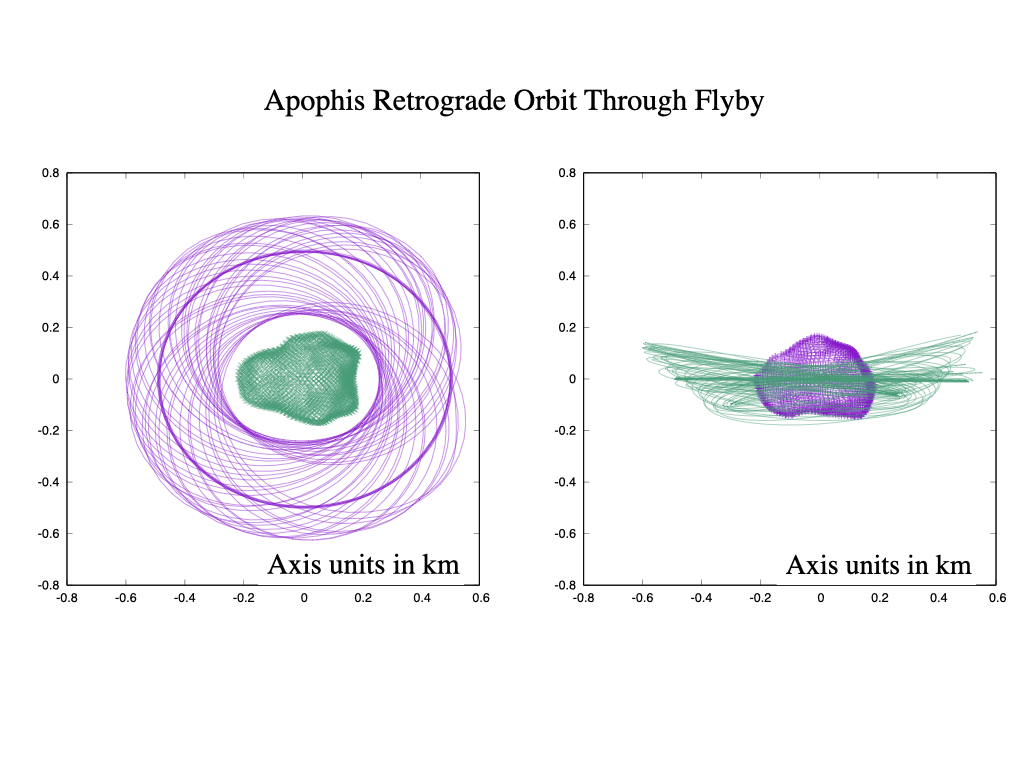}
	\caption{An initially retrograde orbit about Apophis, viewed in an inertial frame. The thick line is the initial orbit prior to flyby. Axis units are in km. }
	\label{fig:retro_orbit}
\end{figure}

\begin{figure}[htb]
	\centering\includegraphics[width=5in]{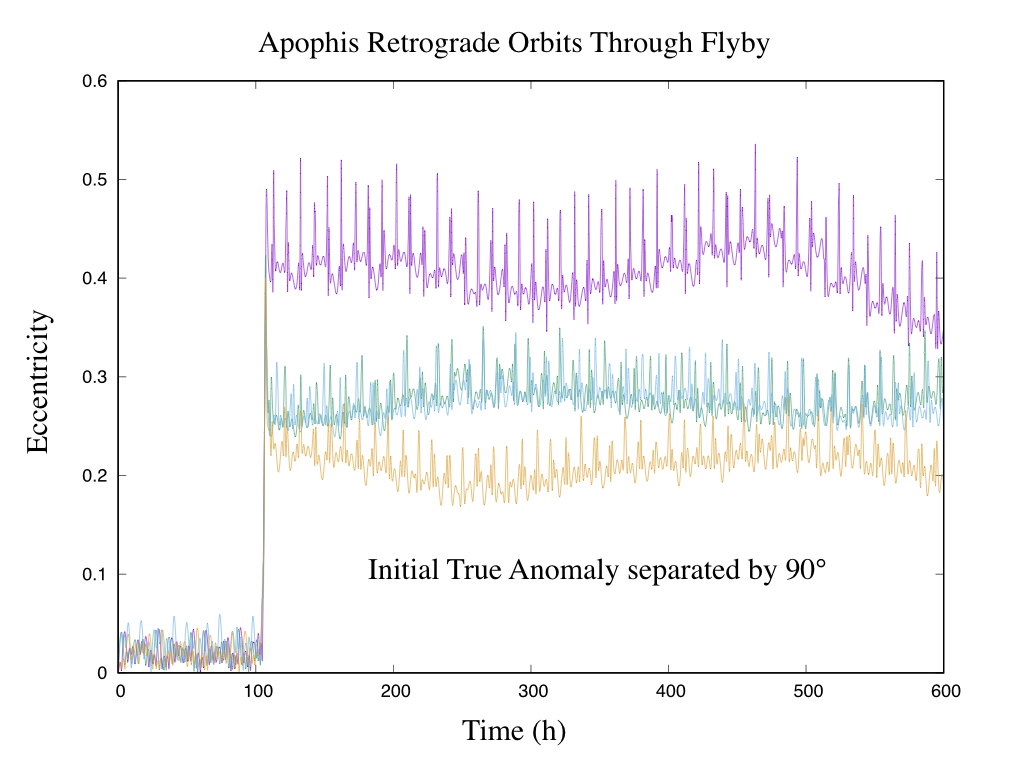}
	\caption{Eccentricities of four initially retrograde orbits started at 90 degree offsets in true anomaly. }
	\label{fig:retro_eccen}
\end{figure}

Finally, we show some examples for initially retrograde orbits about Apophis. These are seen to be more stable in general, and would probably be the most feasible orbit to maintain through flyby. Even still, there is substantial variation in the orbit as a function of its phasing. Here Fig.\ \ref{fig:retro_orbit} shows an example orbit, with a side view showing the extent to which the orbit is given inclination. Figure \ref{fig:retro_eccen} shows the time history of four retrograde orbits given 90$^\circ$ phasing in their initial locations. Again, the initial phasing of the orbit has large impacts on the resulting orbit after flyby.

%Initial investigations show that it is possible for an orbiter to remain trapped in orbit the asteroid, although the orbit will be strongly perturbed across the closest approach. The changes in the orbit elements are sensitive to the orbit orientation relative to the Earth-line, and in some cases a change of node by only 90$^\circ$ can change an orbit from being stable to unstable and either impacting or escaping. Figure \ref{fig:orbits} shows some example trajectories of an orbiter about Apophis as it travels through its Earth close approach. These and other scenarios will be studied in detail. 

\clearpage

\section{Discussion}

This brief and high level analysis considered different approaches to maintaining a spacecraft in proximity of Apophis during its Earth flyby. The analysis was not comprehensive, however basic aspects of the motion were identified and some of the challenges and constraints were discussed. Each of the different modes of operation were seen to be feasible, although the close hovering and orbiting options have some relatively extreme aspects about them. We discuss each of the approaches in the following, and identify what the key issues and challenges may be. 

\paragraph{Distant Dynamics} This approach is perhaps the technically simplest one to apply, as it only requires that the spacecraft be placed relative to the asteroid at the appropriate location, and then the natural dynamics will take the spacecraft past Apophis, ideally along the Earth-Apopohis line at closest approach. One attractive feature of this approach, shown in Fig.\ \ref{fig:flyby} is that the relative distance to Apophis will only vary by a factor of 2 across the range from $f = -\pi/2 \rightarrow \pi/2$. There are no special controls needed to fly in such a trajectory either. 

\paragraph{Hovering Dynamics} The advantage of the hovering dynamics are that they allow the spacecraft to be placed along the Earth-Apophis line throughout the entire flyby (\textcolor{black}{or more generally along any given direction that changes in the Earth-Apophis frame}). These approaches require active control to maintain, although the magnitudes of the control accelerations needed are small enough so that the hovering spacecraft could just provide occasional impulsive $\Delta V$ maneuvers to control the location of the spacecraft, similar to the approach taken for the Hayabusa2 mission. Of the different approaches analyzed, it seems most sensible to implement the fixed distance hovering approach. The need for both lateral and radial control is important to note, however. Finally, if the spacecraft is placed in the equilibrium point it will require no net acceleration, although a closed loop control will still need to be implemented. 

\paragraph{Orbiting Dynamics} Finally, we note that it appears to be feasible to maintain a ballistic orbit through flyby, if appropriately designed. The most robust approach is to fly retrograde, although even these orbits are substantially perturbed. The \textcolor{black}{operations effort} needed to place a spacecraft into a low orbit is significant, however, and requires the spacecraft operations team to navigate precisely, estimate the mass parameters of the asteroid, and be able to precisely phase the spacecraft location in orbit. Even so, the long-term stability of these orbits cannot be guaranteed, and must also be evaluated accounting for solar radiation pressure after the flyby. \textcolor{black}{Thus, even though theoretically feasible, having the spacecraft remain in orbit through the flyby does not seem to be a great candidate approach.}

\paragraph{Effect of Neglected Forces}
\textcolor{black}{
A number of more detailed effects acting on the spacecraft have been neglected in this paper. These include ignoring SRP during the Earth flyby, the use of simple non-gravitational models for the SRP in the heliocentric case, and the neglect of 3rd body perturbations for each of the orbital regimes. Each of these effects can definitely modify the overall dynamics and analysis presented herein, however we believe that  they will only provide secondary effects that can be discounted in these initial analyses. For example, during the close Earth passage the effect of SRP relative to the Earth tidal forces will be small and constant. Previous study has shown that the timescale of SRP effects will be on the order of a few orbit periods about the asteroid, during which time the flyby and its main effects have occurred. Post flyby, the inclusion of SRP for the orbital cases will be important again, however in this paper we do not examine the longer term dynamics of this case. The use of more complex non-gravitational models is difficult to analyze without a specific case study to start from, thus the use of a simple cannonball model is justified. Finally, it is not expected for the moon or sun to play a significant role during the hours when Apophis is close to the Earth, and conversely for the Earth to play a significant role for the heliocentric Apophis trajectory except around the close approach. 
}

\section{Conclusions}

The capability of spacecraft to maintain close proximity to the asteroid Apophis during its close passage to the Earth in 2029 is evaluated. \textcolor{black}{Equations that describe three different operational approaches are presented and analyzed. These include monitoring the asteroid at a distance during the flyby, hovering relative to the asteroid through the flyby and orbiting about the asteroid through the flyby.
We find that each of these approaches can be feasible, although some of them will require more intensive operations and may have a higher risk. Also analyzed is the overall stability of an orbiter relative to Apophis when it is not in close proximity to the Earth. The results presented can be of use in evaluating proximity operations for Apophis as it flies by the Earth, or for any other asteroid or comet that has a close passage to a planetary body. }

\textcolor{black}{
\section*{Acknowledgements} The author acknowledges the very helpful reviews and comments from two referees. Their inputs definitely improved the paper. 
}

\bibliographystyle{AAS_publication}   % Number the references.
\bibliography{references}   % Use references.bib to resolve the labels.

\section*{Appendix}

\textcolor{black}{The following solution has been developed for relative motion about an eccentric orbit \cite{yamanaka2002new,dang2017solutions}.
Here we note that these previous solutions can be extended without loss of generality to the hyperbolic case as well.} 
Define a state vector as $\bar\Xi_0 = \left[ {X}_0, Y_0, Z_0, X_0', Y_0', Z_0' \right]$.  
Then the general orbit solution for linearized motion about an eccentric or hyperbolic orbit can be specified as the linear mapping
\begin{eqnarray}
	\bar\Xi_0 & = & \Phi(f,f_o) \bar\Xi_o 
\end{eqnarray}
where $\Phi \in \mathbf{R}^{6\times6}$ is the state transition matrix for the system. The entries of $\Phi$ can be written out in detail as
\begin{eqnarray}
	\Phi & = & \left[ \begin{array}{cccccc} 
		\phi_{XX} & \phi_{XY} & 0 & \phi_{XX'} & \phi_{XY'} & 0 \\
		\phi_{YX} & \phi_{YY} & 0 & \phi_{YX'} & \phi_{YY'} & 0 \\
		0 & 0 & \phi_{ZZ} & 0 & 0 & \phi_{ZZ'} \\
		\phi_{X'X} & \phi_{X'Y} & 0 & \phi_{X'X'} & \phi_{X'Y'} &  \\
		\phi_{Y'X} & \phi_{Y'Y} & 0 & \phi_{Y'X'} & \phi_{Y'Y'} & 0 \\
		0 & 0 & \phi_{Z'Z} & 0 & 0 & \phi_{Z'Z'} 
			\end{array} \right] \label{eq:STM}
\end{eqnarray}
where we inserted zeros in all of the cross coupling terms between the out-of-plane and in-plane terms. 
The remaining terms are then, taking $f_o = 0$, 
\begin{eqnarray}
	\phi_{XX} & = & \frac{1}{1-e}\left[ 4 + 2e - 3\cos f - 3e\cos^2 f - 3 e (2+e)\sin f(1+e\cos f) L \right]  \\
	\phi_{XY} & = & 0 \\
	\phi_{XX'} & = & \frac{1}{1+e} \sin f (1+e\cos f) \\
	\phi_{XY'} & = & \frac{1}{1-e} \left[ 2 + 2e - 2\cos f - 2e\cos^2 f - 3 e (1+e)\sin f(1+e\cos f) L \right] \\
	\phi_{YX} & = & \frac{1}{1-e} \left[ 3\sin f ( 2 + e \cos f) - 3 (2+e)(1+e\cos f)^2 L \right] \\
	\phi_{YY} & = & 1 \\
	\phi_{YX'} & = & \frac{1}{1+e} \left[ \cos f (2+e\cos f) - (2+e)\right] \\
	\phi_{YY'} & = & \frac{1}{1-e}\left[ 2\sin f (2+e\cos f) - 3(1+e)(1+e\cos f)^2 L \right] \\
	\phi_{X'X} & = & \frac{1}{1-e}\left[ 3\sin f + 3 e \sin2f - 3 e (2+e) ( \cos f + e\cos2f )L  - \frac{3e(2+e)\sin f}{1+e\cos f}\right] \\
	\phi_{X'Y} & = & 0 \\
	\phi_{X'X'} & = & \frac{1}{1+e}\left[ \cos f + e \cos2f\right] \\
	\phi_{X'Y'} & = & \frac{1}{1-e} \left[ 2 \sin f + 2 e \sin2f - 3e (1+e)(\cos f + e\cos2f)L - \frac{3e(1+e)\sin f}{1+e\cos f}\right] \\
	\phi_{Y'X} & = & \frac{1}{1-e} \left[ 6\cos f + 3 e \cos2f - 3(2+e)(1 - 2 e \sin f (1+e\cos f)L)  \right] \\
	\phi_{Y'Y} & = & 0 \\
	\phi_{Y'X'} & = & \frac{-2\sin f(1+e\cos f) }{1+e} \\
	\phi_{Y'Y'} & = & \frac{1}{1-e} \left[ 4 \cos f + 2 e \cos2f - 3(1+e)(1 - 2e\sin f(1+e\cos f)L) \right] \\
	\phi_{ZZ} & = & \cos f \\
	\phi_{ZZ'} & = & \sin f \\
	\phi_{Z'Z} & = & -\sin f \\
	\phi_{Z'Z'} & = & \cos f	
\end{eqnarray}
where we note the function $L$ is defined as
\begin{eqnarray}
	L(f) & = & \int \frac{ df}{(1+e\cos f)^2} \\
	& = & \sqrt{\frac{\mu}{p^3}} t
\end{eqnarray}
where $t$ is the time from perihelion. Thus we see that $L$ will linearly increase in time, and could lead to an overall secular drift.

\end{document}